\documentclass[12pt]{article}

\usepackage{latexsym}
\usepackage{amssymb}
\usepackage{amsthm}
\usepackage{amsmath}
\usepackage{graphicx}

\newcommand{\N}{{\rm I}\!{\rm N}}

\begin{document}

\begin{center}
	{\large
		\bf {A new approach to the modeling of financial volumes}}
\end{center}
\begin{center}
	\begin{small}
		{\sc  Guglielmo D'Amico}\\
		Department of Pharmacy, University "G. d'Annunzio" of Chieti-Pescara\\
		e-mail: \texttt{g.damico@unich.it}\\
		{\sc  Filippo Petroni}\\
		Department of Economy and Business, University of Cagliari, Cagliari\\
	\end{small}
\end{center}

\begin{abstract}
In this paper we study the high frequency dynamic of financial volumes of traded stocks by using a semi-Markov approach.
More precisely we assume that the intraday logarithmic change of volume is described by a weighted-indexed semi-Markov chain model.  Based on this assumptions we show that this model is able to reproduce several empirical facts about volume evolution like time series dependence, intra-daily periodicity and volume asymmetry. Results have been obtained from a real data application to high frequency data from the Italian stock market from first of January 2007 until end of December 2010. 
\end{abstract}

\section{Introduction}
Studies on market microstructure have acquired a crucial importance in order to explain the price formation process, see e.g. De Jong and Rindi (2009). The main variables are (logarithmic) price returns, volumes and duration. Sometimes they are modeled jointly and the main approach is the so-called econometric analysis, see e.g. Manganelli (2005) and Podobnik et al. (2009) and the bibliography therein.\\
\indent The volume variable is very important not only because it interacts directly with duration and returns but also because a correct specification of forecasted volumes can be used for Volume Weighted Average Price trading, see e.g. Brownlees et al. (2011). This variable has been investigated for long time and several statistical regularities have been highlighted, see e.g. Jain and Joh (1988). \\
\indent In this paper we propose an alternative approach to the modeling of financial volumes which is based on a generalization of semi-Markov processes called Weighted-Indexed Semi-Markov Chain (WISMC) model.  This choice is motivated by recent results in the modeling of price returns in high-frequency financial data where the WISMC approach was demonstrated to be particular efficient in reproducing the statistical properties of financial returns, see D'Amico and Petroni (2011, 2012, 2014). The WISMC model is very flexible and for this reason we decided to test its appropriateness also for financial volumes.\\  
\indent It should be remarked that WISMC models generalize semi-Markov processes and also non-Markovian models based on continuous time random walks that were used extensively in the econophysics community, see e.g. Mainardi et al. (2000) and Raberto et al. (2002).\\
\indent The model is applied to a database of high frequency volume data from all the stocks in the Italian Stock Market from first of January 2007 until end of December 2010. In the empirical analysis we find that WISMC model is a good choice for modeling financial volumes which is able to correctly reproduce stylized facts documented in literature of volumes such as the autocorrelation function.\\
\indent The paper is divided as follows: First, Weighted-Indexed-Semi-Markov chains are shortly described in Section 2. Next, we introduce our model of financial volumes and an application to real high frequency data illustrates the results. Section 4 concludes and suggests new directions for future developments.

\section{Weighted-Indexed Semi-Markov Chains}

The general formulation of the WISMC as developed in D'Amico and Petroni (2012, 2014) is here only discussed informally. \\
\indent WISMC models share similar ideas as those that generate Markov processes and semi-Markov processes. These processes are all described by
a set of finite states $J_n$ whose transitions are ruled by a transition probability matrix. The semi-Markov process differs from
the Markov process because the transition times $T_n$ where the states change values, are generated according to random variables. Indeed, the time between
transitions $T_{n+1} - T_{n}$ is random and may be modeled by means of any type of distribution functions. In order to better represent the statistical characteristics of high-frequency financial data, in a recent article, the idea of a WISMC model was introduced in the field of price returns, see D'Amico and Petroni (2012). The novelty, with respect to the semi-Markov case, consists in the introduction of a third random variable defined as follows:
\begin{equation}
\label{funcrela}
I_{n}(\lambda)=\sum_{k=0}^{n-1}\sum_{a=T_{n-1-k}}^{T_{n-k}-1}f^{\lambda}(J_{n-1-k},T_{n},a),
\end{equation}
where $f$ is any real value bounded function and $I_{0}^{\lambda}$ is known and non-random. The variable $I_{n}^{\lambda}$ is designated to summarize the information contained in the past trajectory of the $\{J_{n}\}$ process that is relevant for future predictions. Indeed, at each past state $J_{n-1-k}$ occurred at time $a\in \N$ is associated the value $f^{\lambda}(J_{n-1-k},T_{n},a)$, which depends also on the current time $T_{n}$. The quantity $\lambda$ denotes a parameter that represents a weight and should be calibrated to data. In the applicative section we will describe the calibration of $\lambda$ as well as the choice of the function $f$.\\
\indent The WISMC model is specified once a dependence structure between the variables is considered. Toward this end, the following assumption is formulated:
\begin{equation}
\label{kernel}
\begin{aligned}
& \mathbb{P}[J_{n+1}=j,\: T_{n+1}-T_{n}\leq t |J_{n},T_{n},I_{n}^{\lambda},J_{n-1},T_{n-1},I_{n-1}^{\lambda},\ldots]\\
& =\mathbb{P}[J_{n+1}=j,\: T_{n+1}-T_{n} \leq t |J_{n},I_{n}^{\lambda}]:=Q_{J_{n}j}^{\lambda}(I_{n}^{\lambda};t).
\end{aligned}
\end{equation}
\indent Relation $(\ref{kernel})$ asserts that the knowledge of the values of the variables $J_{n}, I_{n}^{\lambda}$ is sufficient to give the conditional distribution of the couple $J_{n+1}, T_{n+1} - T_{n}$ whatever the values of the past variables might be. Therefore to make probabilistic forecasting we need the knowledge of the last state of the system and the last value of the index process. If ${\bf Q}^{\lambda}(x;t)$ is constant in $x$ then the WISMC kernel degenerates in an ordinary semi-Markov kernel and the WISMC model becomes equivalent to classical semi-Markov chain models, see e.g. D'Amico and Petroni (2012a) and Fodra and Pham (2015).\\
\indent The probabilities $Q_{ij}^{\lambda}(x;t))_{i,j\in E}$ can be estimated directly using real data. In D'Amico and Petroni (2017) it is shown that the estimator
\[
\hat{Q}_{i,j}^{\lambda}(x;t):=\frac{N_{ij}(x;t)}{N_{i}(x)},
\]
is the approached maximum likelihood estimator of the corresponding transition probabilities. The quantity $N_{ij}(x;t)$ expresses the number of transitions from state $i$, with an index value $x$, to state $j$ with a sojourn time in state $i$ less or equal to $t$. The quantity $N_{i}(x)$ is the number of visits to state $i$ with an index value $x$.

\section{The volume model}\label{volume}

\indent Let us assume that the trading volume of the asset under study is described by the time varying process $V(t)$, $t\in \N$.\\
\indent The (logarithmic) change of volume at time $t$ over the unitary time interval is defined by
\begin{equation}
Z^{V}(t) = \log \frac{V(t + 1)}{V(t)}.
\end{equation}
On a short time scale, $Z^{V}(t)$ changes value in correspondence of an increasing sequence of random times, $\{T_{n}^{V}\}_{n\in \N}$. According to the notation adopted in the previous section, we denote the values assumed at time $T_{n}^{V}$ by $J_{n}^{V}$ and the corresponding values of the index process by
\begin{equation}\label{funcrela}
I_{n}^{V}(\lambda)=\sum_{k=0}^{n-1}\sum_{a=T_{n-1-k}^{V}}^{T_{n-k}^{V}-1}f^{\lambda}(J_{n-1-k}^{V},T_{n}^{V},a).
\end{equation}
\indent If we assume that the variables $(J_{n}^{V},T_{n}^{V},I_{n}^{V}(\lambda))$ satisfy relationship $(\ref{kernel})$ then the volume process can be described by a WISMC model and if $N^{V}(t)=\sup\{n\in \N: T_{n}^{V}\leq t\}$ is the number of transition of the volume process, then $Z^{V}(t)=J_{N^{V}(t)}^{V}$ is the WISMC process that describes the volume values at any time $t$.\\
\indent We assume also that the set $E$ is finite and is obtained by an opportune discretization of the values of the financial volumes. A description of the adopted discretization and of the state space model is described in next section. The choice of a finite state space can be formalized by defining the state space by
$$E=\{-z_{min}\Delta,\ldots,-2\Delta,-\Delta,0,\Delta,2\Delta,\ldots, z_{max}\Delta\}.$$

\indent Our objective is to demonstrate that a WISMC model for financial volume is able to reproduce some important facts of volume dynamics. One of the most important feature is the persistence of the volume process. As found in Manganelli (2005), the volume process is strongly persistent for frequently traded stocks. A possible explanations is that the volume is pushed up by the arrival of new information to the market participants and some times is necessary in order to delete the effects of the information arrival. In order to investigate the goodness of the WISMC model we define the autocorrelation of the modulus of volumes as 
\begin{equation}
\label{autosquare}
\Sigma(t, t+\tau)=Cov(\mid Z^{V}(t+\tau)\mid,\mid Z^{V}(t)\mid).
\end{equation}
\indent Another important statistic is the first passage time distribution of the volume process. First we need to define the accumulation factor of the volume process from the generic time $t$ to time $t+\tau$:
\begin{equation}
M_{t}^{V}(\tau) = \frac{V(t+\tau)}{V(t)}=e^{\sum_{r=0}^{\tau -1}Z^{V}(t+r)}.
\end{equation}
We will denote the fpt by 
$$
\Gamma_\sigma=\min\{\tau \geq 0 ; M_{0}^{V}(\tau) \geq \sigma\},
$$
where $\sigma$ is a given threshold. We are interested in finding the distributional properties of the fpt, that is to compute  
$$
\mathbb{P}[\Gamma_{\sigma} >t|(J^{V},T^{V},I^{V}(\lambda))_{-m}^{0}=(i,t)_{-m}^{0}],
$$
where
\begin{equation*}
({\bf{J}^{V}})_{-m}^{0}=({J}_{-m}^{V},{J}_{-m+1}^{V},\ldots, {J}_{0}^{V}),
\end{equation*}
\begin{equation*}
({\bf{T}^{V}})_{-m}^{0}=({T}_{-m}^{V},{T}_{-m+1}^{V},\ldots, {T}_{0}^{V}),
\end{equation*}
\begin{equation*}
({\bf{I}^{V}}(\lambda))_{-m}^{0}=({I}_{-m}^{V}(\lambda),{I}_{-m+1}^{V}(\lambda),\ldots, {I}_{0}^{V}(\lambda)).
\end{equation*}

\section{Application to real high frequency data}

The data used in this work are tick-by-tick quotes of indexes and stocks downloaded 
from $www.borsaitaliana.it$ for the period January 2007-December 2010 (4 full years). 
The data have been re-sampled to have 1 minute frequency. Every minutes the last price and
the cumulated volume (number of transaction) is recorded. For each stock the database
is composed of about $5*10^5$ volumes and prices. The list of stocks analyzed and their symbols are reported in Table \ref{tab}. 
\begin{table}
	\begin{center}
		\begin{tabular}{|l|l|}
			\hline			
			\textbf{F} & Fiat\\\hline	
			\textbf{ISP} & Intesa San Paolo\\\hline
			\textbf{TIT} & Telecom\\\hline
			\textbf{TEN} & Tenaris\\\hline
		\end{tabular}
	\end{center}
	\caption{Stocks used in the application and their symbols}\label{tab}
\end{table}

\begin{figure}
	\centering
	\includegraphics[height=10cm]{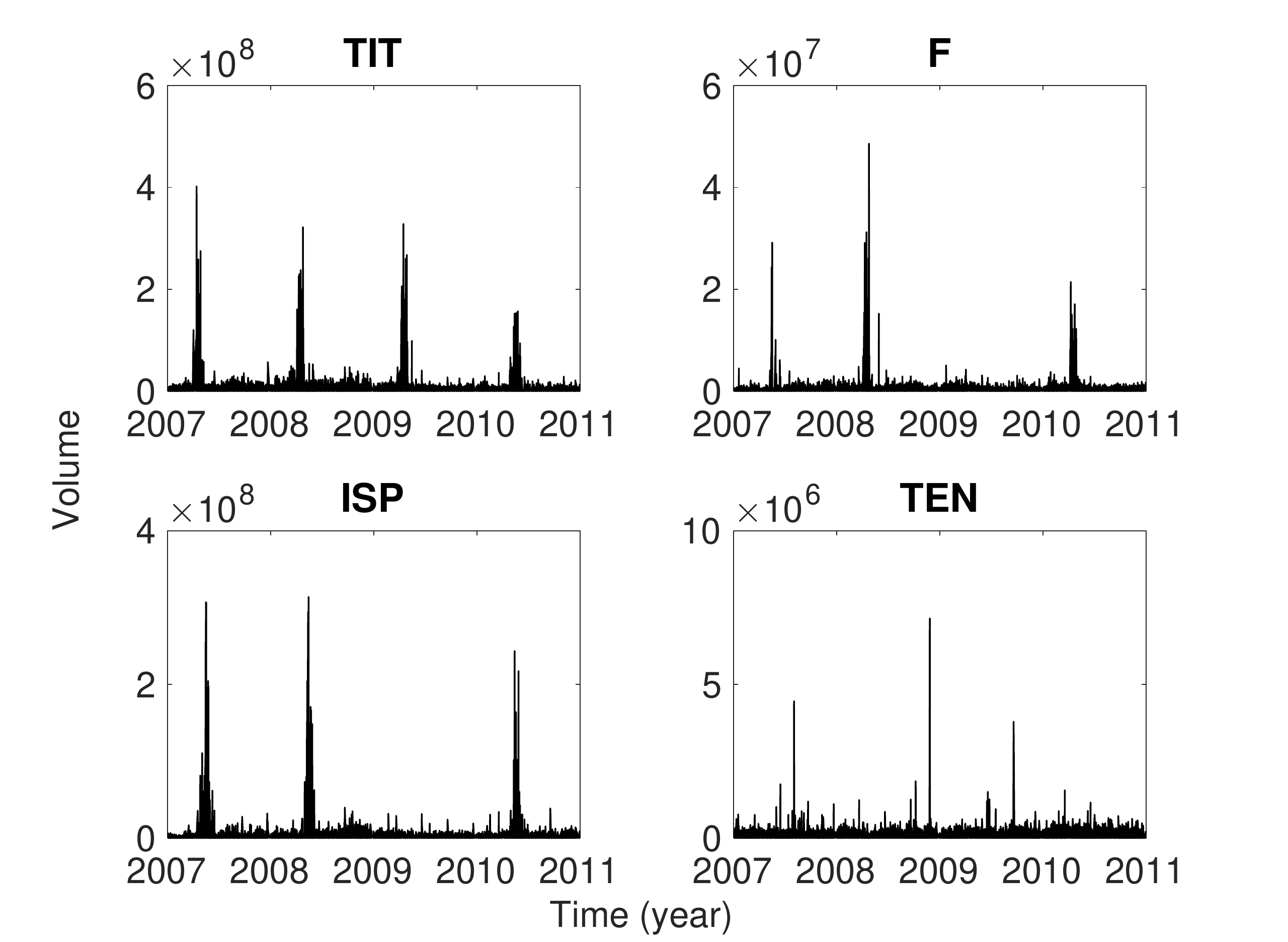}
	\caption{Trading volume of the analyzed stocks.}\label{fig1}
\end{figure}

\begin{figure}
	\centering
	\includegraphics[height=10cm]{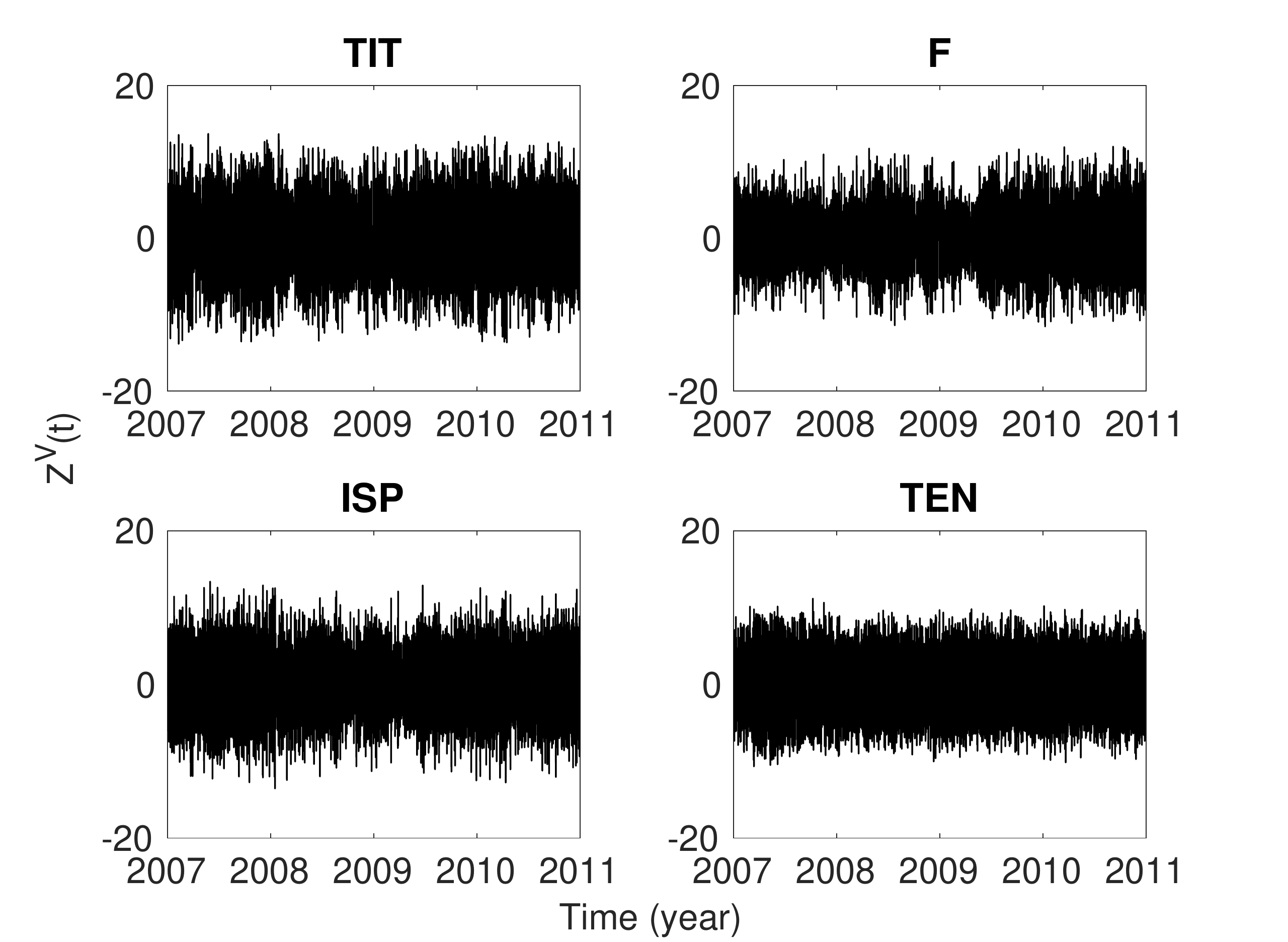}
	\caption{The (logarithmic) change of trading volume of the analyzed stocks.}\label{fig2}
\end{figure}

In Figures \ref{fig1} and \ref{fig2} are shown the trading volume and the logarithmic change of the trading volume for the 4 stocks in the analyzed period.

\begin{figure}
	\centering
	\includegraphics[height=10cm]{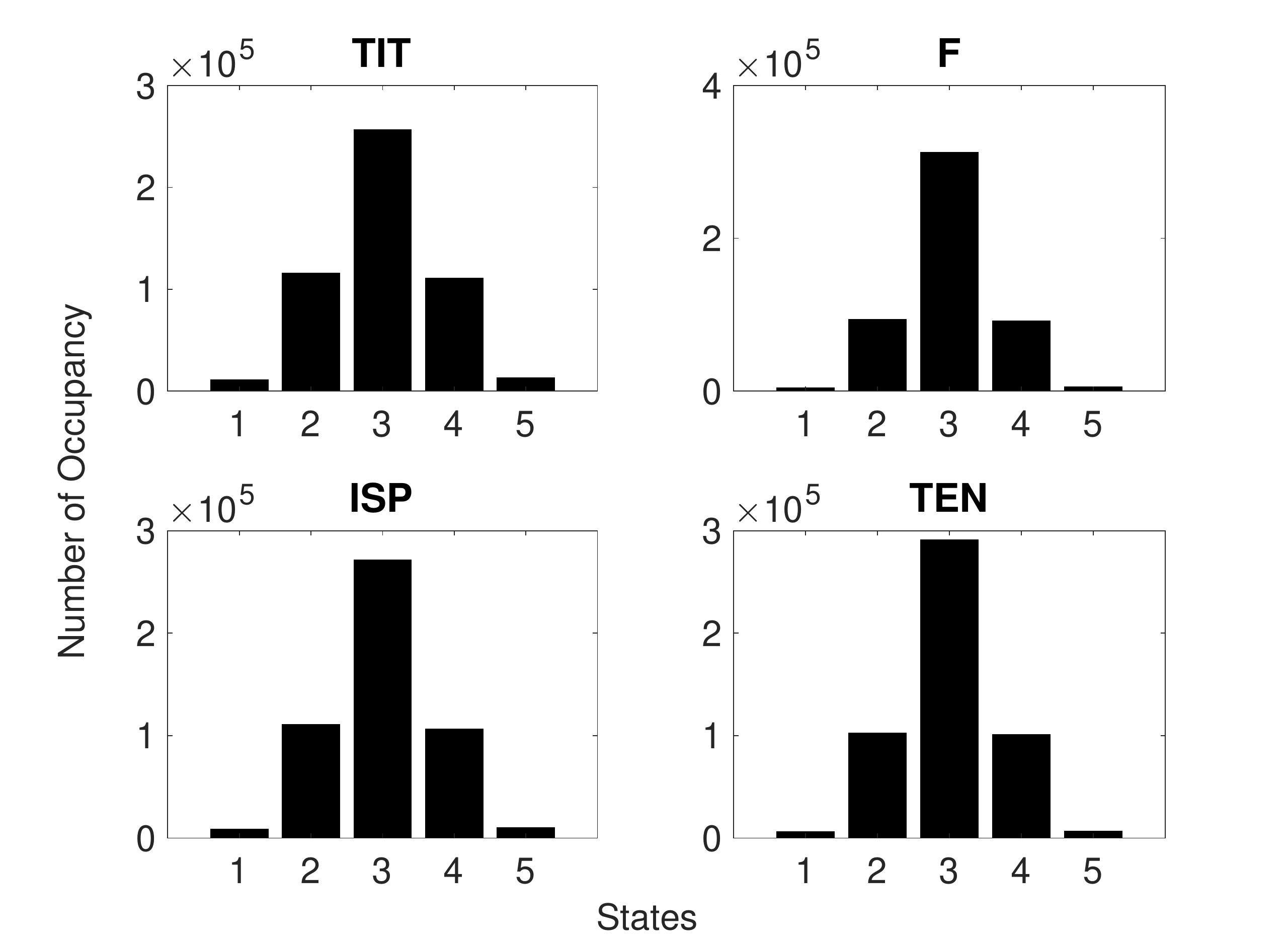}
	\caption{Number of occupancy of each discretized state of the logarithm of volume change.}\label{fig3}
\end{figure}

To model $Z^{V}(t)$ as a WISMC model the first step is to discretize the variable. We choose 5 states of discretization with the following edges: $\{-\infty,-4,-1,1,4,\infty\}$. The number of times that  $Z^{V}(t)$ fall in each state is shown in Figure \ref{fig3}.

Following D'Amico $\&$ Petroni (2012b) we use as definition of the function $f^{\lambda}$ in  (\ref{funcrela}) an exponentially weighted moving average (EWMA) of the squares of $Z^{V}(t)$ which has the following expression:
\begin{equation}
\label{funct}
f^{\lambda}(J_{n-1-k},T_{n},a)=\frac{\lambda^{T_{n}-a} J_{n-1-k}^2}{\sum_{k=0}^{n-1}\sum_{a=T_{n-1-k}}^{T_{n-k}-1}\lambda^{T_{n}-a}}=\frac{\lambda^{T_{n}-a} J_{n-1-k}^2}{\sum_{a=1}^{T_{n}}\lambda^{a}}
\end{equation}
\noindent and consequently the index process becomes
\begin{equation}
\label{ewma}
I_{n}^{V}(\lambda)=\sum_{k=0}^{n-1}\sum_{a=T_{n-1-k}}^{T_{n-k}-1}\Bigg(\frac{\lambda^{T_{n}-a} J_{n-1-k}^2}{\sum_{a=1}^{T_{n}}\lambda^{a}}\Bigg).
\end{equation}

The index $I_{n}^{V}(\lambda)$ was also discretized into 5 states of low, medium low, medium, medium high and high volume variation.
Using these definitions and discretizations we estimated, for each stock, the probabilities defined in the previous section by using their estimators directly from real data. 
By means of Monte Carlo simulations we were able to produce, for each of the 4 stocks, a synthetic time series. 

Each time series is a realization of the stochastic process described in the previous section with the same time length as real data.
Statistical features of these synthetic time series are then compared with the statistical features of real data. In particular, we tested our model for the ability to reproduce the autocorrelation functions of the absolute value of $Z^{V}(t)$ and the first passage time distribution.

\begin{figure}
	\centering
	\includegraphics[height=10cm]{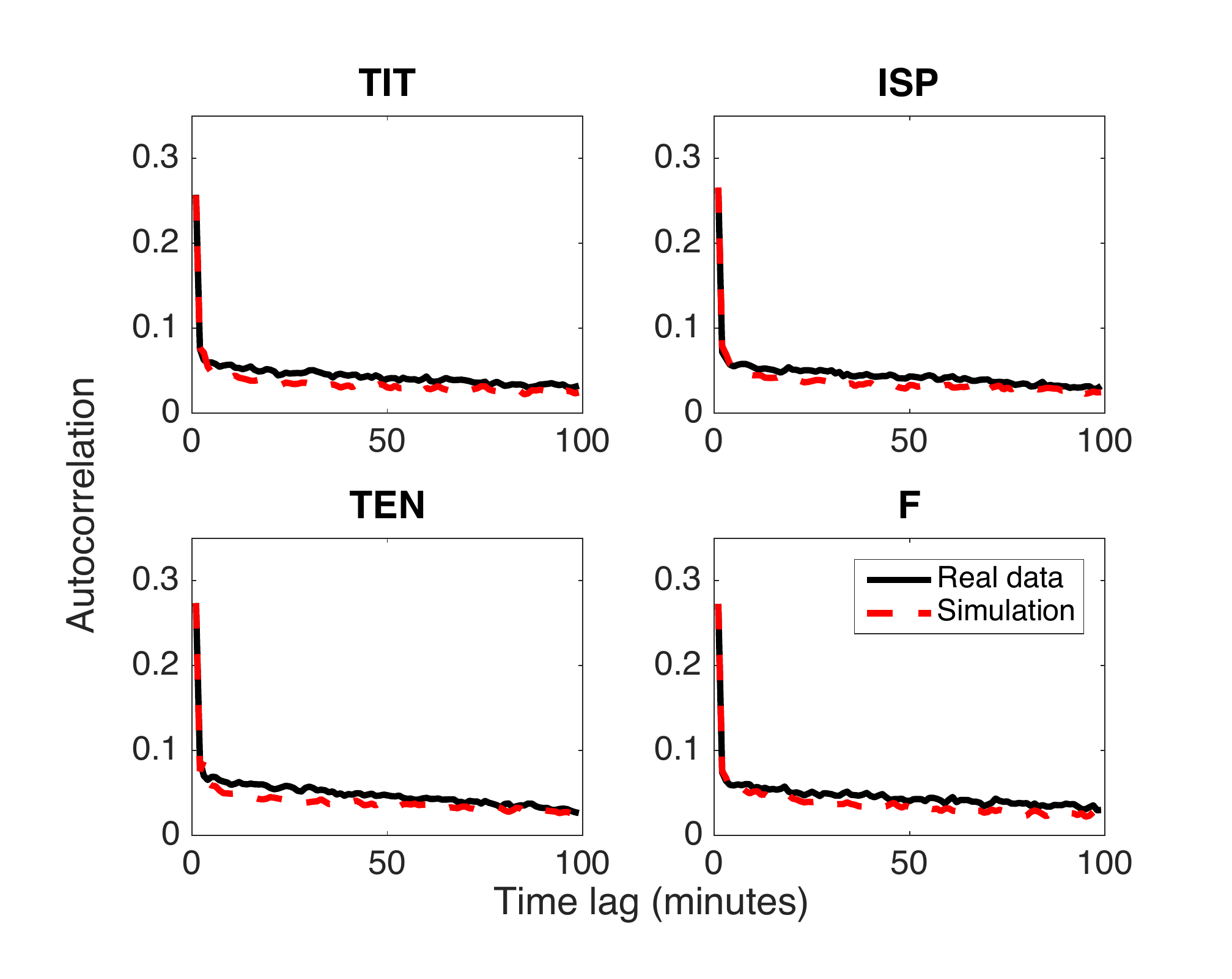}
	\caption{Autocorrelation functions of the absolute value of $Z^{V}(t)$ for real data (solid line) and synthetic (dashed line) time series.}\label{fig4}
\end{figure}

We estimated $\Sigma(\tau)$ (see equation \ref{autosquare}) for real data and for synthetic data and show in Figure \ref{fig4} a comparison between them for all stocks. For each stock we estimated the percentage root mean square error (RMSE) the results are reported in Table \ref{table_corr}. 

\begin{table}
	\begin{center}
		\begin{tabular}{|l|l|}
			\hline\noalign{\smallskip}
			\textbf{Stock} & \textbf{Error}\\
			\noalign{\smallskip}\hline
			\textbf{F}&$3.6\%$\\\hline	
			\textbf{ISP}&$3.2\%$\\\hline
			\textbf{TIT}&$3.0\%$\\\hline
			\textbf{TEN}&$3.9\%$\\\hline
		\end{tabular}
	\end{center}
	\caption{Percentage square root mean error between real and synthetic autocorrelation function reported in Figure \ref{fig4}.}\label{table_corr}
\end{table}

It is possible to note that our model is able to reproduce almost perfectly the autocorrelation of the absolute value of the logarithmic volume change stocks. 

For each of the stocks in our database we estimate the first passage time distribution directly from the data (real data) and from the synthetic time series generated as described above.\\

\begin{figure}
\centering
\includegraphics[width=8cm]{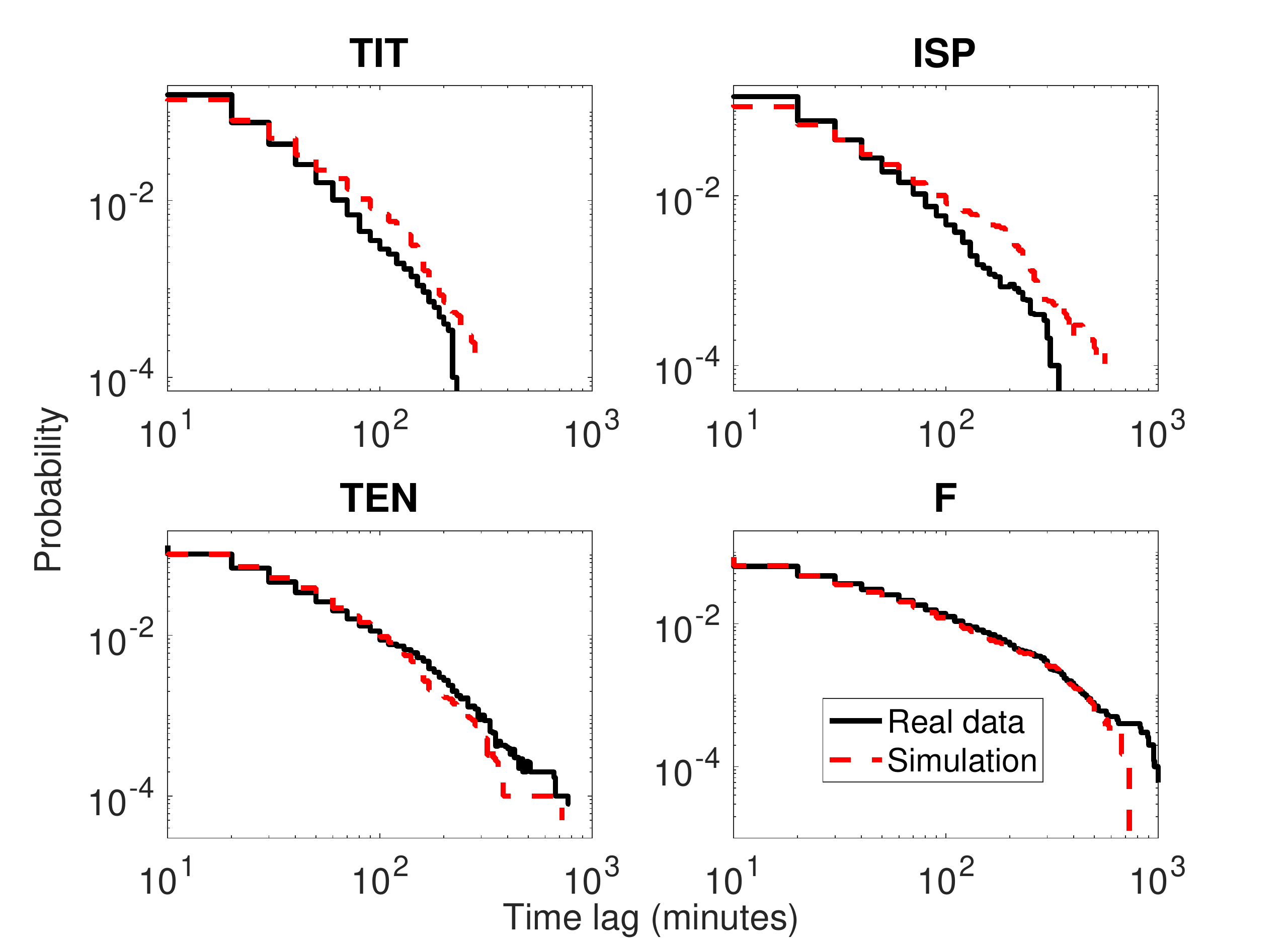}
\caption{First passage time distribution for $\rho=1000$} \label{fpt}
\end{figure}

From Figure \ref{fpt} it is obvious that the WISMC model is also able to reproduce the first passage time distribution of real data.

\section{Conclusions}
In this paper we advanced the use of Weigthed-Indexed Semi-Markov Chain models for modeling high frequency financial volumes. We applied the model on real financial data and we shown that the model is able to reproduce important statistical fact of financial volumes as the autocorrelation function of the absolute values of volumes and the first passage time distribution. Further developments will be a more extensive application to other financial stocks and indexes and the proposal of a complete model where returns, volumes and durations are jointly described.


\begin{thebibliography}{99}

\bibitem{brow11} C.T. Brownlees, F. Cipollini, G.M. Gallo. Journal of Financial Econometrics 9(3) (2011) 489-518.
\bibitem{dape11} G. D'Amico, F. Petroni, "A semi-Markov model with memory for price changes", Journal of Statistical Mechanics: Theory and Experiment, P12009 (2011).
\bibitem{dape12} G. D'Amico, F. Petroni, "Weighted-indexed semi-Markov models for modeling financial returns", Journal of Statistical Mechanics: Theory and Experiment, P07015 (2012).
\bibitem{dape12} G. D'Amico, F. Petroni, "A semi-Markov model for price returns", Physica A 391 (2012) 4867-4876.
\bibitem{dape14} G. D'Amico, F. Petroni, "Multivariate high-frequency financial data via semi-Markov processes", Markov Processes and Related Fields, 20 (2014) 415-434.
\bibitem{dape17} G. D'Amico, F. Petroni, "Copula Based Multivariate Semi-Markov Models with Applications in High-Frequency Finance", Submitted.
\bibitem{dejo09} F. De Jong, B. Rindi, "The Microstructure of Financial Markets", Cambridge University Press (2009), Cambridge, New York.
\bibitem{jain88} Jain, P., Joh, G., 1988. "The dependence between hourly prices and trading volume", Journal of
Financial and Quantitative Analysis 23, 269-283.
\bibitem{mai00} F. Mainardi, M. Raberto, R. Gorenflo, E. Scalas. Physica A 287 (2000) 468.
\bibitem{mang05} S. Manganelli, "Duration, Volume and Volatility impact of trades", Journal of Financial Markets, 8, (2005) 377-399.
\bibitem{podo09} B. Podobnik, D. Horvatic, A.M. Petersen, H.E. Stanley, "Cross-Correlations between Volume Change and price Change", PNAS, December 29(52) (2009) 22079-22084.
\bibitem{rab02} M. Raberto, E. Scalas, F. Mainardi. Physica A 314 (2002) 749.
\end{thebibliography}
\end{document}